\documentclass[aps,prl,superscriptaddress,amsmath,amssymb,preprint]{revtex4-1}

\usepackage[utf8x]{inputenc}
\usepackage[english]{babel}
\usepackage{amsmath}
\usepackage{amssymb}

\usepackage{amstext}
\usepackage{graphicx}
\usepackage{lmodern}
\usepackage{float}
\usepackage{siunitx}
\usepackage{color}

\usepackage[colorlinks=true]{hyperref}

\begin{document}

\title{Complex THz and DC inverse spin Hall effect in YIG/Cu$_{1-x}$Ir$_{x}$  bilayers across a wide concentration range}

\author{Joel Cramer}
\affiliation{Institute of Physics, Johannes Gutenberg-University Mainz, 55099 Mainz, Germany}
\affiliation{Graduate School of Excellence Materials Science in Mainz, 55128 Mainz, Germany}

\author{Tom Seifert}
\affiliation{Department of Physical Chemistry, Fritz Haber Institute of the Max Planck Society,
14195 Berlin, Germany}

\author{Alexander Kronenberg}
\affiliation{Institute of Physics, Johannes Gutenberg-University Mainz, 55099 Mainz, Germany}

\author{Felix Fuhrmann}
\affiliation{Institute of Physics, Johannes Gutenberg-University Mainz, 55099 Mainz, Germany}

\author{Gerhard Jakob}
\affiliation{Institute of Physics, Johannes Gutenberg-University Mainz, 55099 Mainz, Germany}

\author{Martin Jourdan}
\affiliation{Institute of Physics, Johannes Gutenberg-University Mainz, 55099 Mainz, Germany}

\author{Tobias Kampfrath}
\affiliation{Department of Physical Chemistry, Fritz Haber Institute of the Max Planck Society,
14195 Berlin, Germany} \affiliation{Department of Physics, Freie Universit{\"a}t Berlin, 14195
Berlin, Germany}

\author{Mathias Kläui}
\email{Klaeui@uni-mainz.de} \affiliation{Institute of Physics, Johannes Gutenberg-University Mainz,
55099 Mainz, Germany} \affiliation{Graduate School of Excellence Materials Science in Mainz, 55128
Mainz, Germany}

\date{\today}

\begin{abstract}	

We measure the inverse spin Hall effect of Cu$_{1-x}$Ir$_{x}$ thin films on yttrium iron garnet over a wide range of Ir concentrations ($0.05 \leqslant x \leqslant 0.7$).
Spin currents are triggered through the spin Seebeck effect, either by a DC temperature gradient or by ultrafast optical heating of the metal layer.
The spin Hall current is detected by, respectively, electrical contacts or measurement of the emitted THz radiation.
With both approaches, we reveal the same Ir concentration dependence that follows a novel complex, non-monotonous behavior as compared to previous studies.
For small Ir concentrations a signal minimum is observed, while a pronounced maximum appears near the equiatomic composition.
We identify this behavior as originating from the interplay of different spin Hall mechanisms as well as a concentration-dependent variation of the integrated spin current density in Cu$_{1-x}$Ir$_{x}$.
The coinciding results obtained for DC and ultrafast stimuli show that the studied material allows for efficient spin-to-charge conversion even on ultrafast timescales, thus enabling a transfer of established spintronic measurement schemes into the terahertz regime.

\end{abstract}

\maketitle

\section{Introduction}
Spin currents are a promising ingredient for the implementation of next-generation, energy-efficient spintronic applications.
Instead of exploiting the electronic charge, transfer as well as processing of information is mediated by spin angular momentum.
Crucial steps towards the realization of spintronic devices are the efficient generation, manipulation and detection of spin currents at highest speeds possible.
Here, the spin Hall effect (SHE) and its inverse (ISHE) are in the focus of current research \cite{Sinova2015} as they allow for an interconversion of spin and charge currents in heavy
metals with strong spin-orbit interaction (SOI).
The efficiency of this conversion is quantified by the spin Hall angle $\theta_{\mathrm{SH}}$.

In general, the SHE has intrinsic as well as extrinsic spin-dependent contributions.
The intrinsic SHE results from a momentum-space Berry phase effect and can, amongst others, be observed in 4$d$ and 5$d$ transition metals \cite{Tanaka2008,Morota2011,Sinova2015}.
The extrinsic SHE, on the other hand, is a consequence of skew and side-jump scattering off impurities or defects \cite{Fert2011}.
It occurs in (dilute) alloys of normal metals with strong SOI impurity scatterers \cite{Niimi2011,Niimi2012,Zou2016, Ramaswamy2017}, but can also be prominent in pure metals in the superclean regime \cite{Sagasta2016}.
As a consequence, the type of employed metals and the alloy composition are handles to adjust and maximize the SHE.
Remarkably, it was recently shown that the SHE in alloys of two heavy metals (e.g. AuPt) can even exceed the SHE observed for the single alloy
partners \cite{Obstbaum2016}.
Pioneering work within this research field covered the extrinsic SHE by skew scattering in copper-iridium alloys \cite{Niimi2011}.
However, previously the iridium concentration was limited to \SI{12}{\percent} effective doping of Cu with dilute Ir.
The evolution of the SHE in the alloy regime for large concentration thus remains an open question and the achievable maximum by an optimized alloying strategy is unknown.

The potential of a metal for spintronic applications (i.e. $\theta_{\mathrm{SH}}$) can be quantified by injecting a spin current and measuring the resulting charge response.
This can be accomplished by, for instance, coherent spin pumping through ferromagnetic resonance \cite{Tserkovnyak2002,Mizukami2002,Saitoh2006} or the spin Seebeck effect (SSE) \cite{Bauer2012,Uchida2016}.
The SSE describes the generation of a magnon spin current along a temperature gradient within a magnetic material.
Typically, such experiments involve a heterostructure composed of a magnetic insulator, such as yttrium iron garnet (YIG), and the ISHE metal under study [see Fig. \ref{fig:experiment}(a)].
A DC temperature gradient in the YIG bulk is induced by heating the sample from one side.
On the femtosecond timescale, however, a temperature difference and thus a spin current across the YIG-metal interface can be induced by heating the metal layer with an optical laser pulse [Fig.~\ref{fig:experiment}(b)] \cite{Agrawal2014,schreier2016spin,Kimling2016,Seifert}.
This interfacial SSE has been shown to dominate the spin current in the metal on timescales below $\sim 300$~ns \cite{Agrawal2014}. 

For ultrafast laser excitation, the resulting sub-picosecond ISHE current leads to the emission of electromagnetic pulses at frequencies extending into the terahertz (THz) range, which can be detected by optical means \cite{kampfrath2013terahertz}.
Therefore, femtosecond laser excitation offers the remarkable benefit of contact-free measurements of the ISHE current without any need of micro-structuring the sample.
The all-optical generation as well as detection of ultrafast electron spin currents \cite{kampfrath2013terahertz,Seifert2016a} is a key requirement for transferring spintronic concepts into the THz range \cite{Walowski2016}.
So far, however, characterization of the ISHE was conducted by experiments including DC spin current signals as, for instance, the bulk SSE [Fig.~\ref{fig:experiment}(a)].
For the use in ultrafast applications, it thus remains to be shown whether alloying yields the same notable changes of the spin-to-charge conversion efficiency in THz interfacial SSE experiments [Fig.~\ref{fig:experiment}(b)] and whether alloys can provide an efficient spin-to-charge conversion even at the ultrafast timescale.

In this work, we study the compositional dependence of the ISHE in YIG/Cu$_{1-x}$Ir$_{x}$ bilayers over a wide concentration range ($0.05 \leqslant x \leqslant 0.7$), exceeding the dilute doping phase investigated in previous studies \cite{Niimi2011}.
The ISHE response of Cu$_{1-x}$Ir$_{x}$ is measured as a function of $x$, for which both DC bulk and THz interfacial SSE are employed.
Eventually, we compare the spin-to-charge conversion efficiency in the two highly distinct regimes of DC and terahertz dynamics across a wide alloying range.
\section{Experiment}
The YIG samples used for this study are of \SI{870}{\nano\meter} thickness, grown epitaxially on (111)-oriented Gd\textsubscript{3}Ga\textsubscript{5}O\textsubscript{12} (GGG) substrates by liquid-phase-epitaxy.
After cleaving the GGG/YIG into samples of dimension \SI{2.5 x 10 x 0.5}{\milli\meter}, Cu$_{1-x}$Ir$_{x}$ thin films (thickness $d_{\mathrm{CuIr}} = \SI{4}{\nano\meter}$) of varying composition ($x = 0.05, 0.1, 0.2, 0.3, 0.5$ and 0.7) are deposited by multi-source magnetron sputtering.
To prevent oxidation of the metal film, a \SI{3}{\nano\meter} Al capping layer is deposited, which, when exposed to air, forms an AlO$_{x}$ protection layer.
For the contact-free ultrafast SSE measurements, patterning of the Cu$_{1-x}$Ir$_{x}$ films into defined nanostructures is not necessary.
In the case of DC SSE measurements, the unpatterned film is contacted for the detection of the thermal voltage.

\begin{figure}[!tb]
	\centering
	\includegraphics[width = 62 mm]{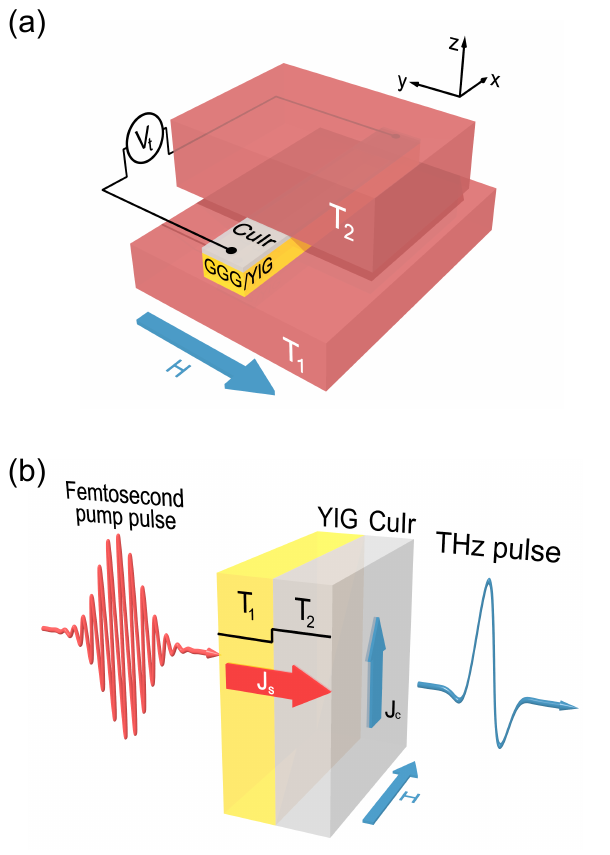}
	\caption{
		(a) Scheme of the setup used for DC SSE measurements. The out-of-plane temperature gradient is
		generated by two copper blocks set to individual temperatures $T_1$ and $T_2$. An external magnetic
		field is applied in the sample plane. The resulting thermovoltage $V_{\mathrm{t}}$ is recorded by a nanovoltmeter.
		(b) Scheme of the contact-free ultrafast SSE/ISHE THz emission approach. The in-plane magnetized
		sample is illuminated by a femtosecond laser pulse, inducing a step-like temperature gradient
		across the YIG/Cu$_{1-x}$Ir$_{x}$ interface. The SSE-induced THz spin current in the CuIr layer is
		subsequently converted into a sub-picosecond in-plane charge current by the ISHE, thereby leading
		to the emission of a THz electromagnetic pulse into the optical far-field. }
	\label{fig:experiment}
\end{figure}

The DC SSE measurements are performed at room temperature in the conventional longitudinal
configuration \cite{Uchida2016}. While an external magnetic field is applied in the sample plane,
two copper blocks, which can be set to individual temperatures, generate a static out-of-plane
temperature gradient, see Fig.~\ref{fig:experiment}(a). This thermal perturbation results in a magnonic
spin current in the YIG layer \cite{Kehlberger2015}, thereby transferring angular momentum into the
Cu$_{1-x}$Ir$_{x}$. A spin accumulation builds up, diffuses as a pure spin current and is
eventually converted into a transverse charge current by means of the ISHE, yielding a measurable
voltage signal. The spin current and consequently the thermal voltage change sign when the YIG
magnetization is reversed. The SSE voltage $V$\textsubscript{SSE} is defined as the difference
between the voltage signals obtained for positive and negative magnetic field divided by 2. Since
$V$\textsubscript{SSE} is the result of the continuous conversion of a steady spin current, it can,
applying the notation of conventional electronics, be considered as a DC signal.

For the THz SSE measurements, the same in-plane magnetized YIG/Cu$_{1-x}$Ir$_{x}$ samples are illuminated at room
temperature by femtosecond laser pulses (energy of \SI{2.5}{\nano\joule}, duration of
\SI{10}{\femto\second}, center wavelength of $\SI{800}{\nano\meter}$ corresponding to a photon
energy of $\SI{1.55}{\electronvolt}$, repetition rate of \SI{80}{\mega\hertz}) of a Ti:sapphire
laser oscillator. Owing to its large bandgap of $\SI{2.6}{\electronvolt}$ \cite{Metselaar1974}, YIG
is transparent for these laser pulses. They are, however, partially (about \SI{50}{\percent}) absorbed by the electrons of
the Cu$_{1-x}$Ir$_{x}$ layer.
The spatially step-like temperature gradient across the YIG/metal interface leads to an ultrafast spin current in the metal
layer polarized parallel to the sample magnetization \cite{Seifert}.
Subsequently, this spin current is converted into a transverse sub-picosecond charge current through the ISHE, resulting in the emission of a THz electromagnetic pulse into the optical far-field. The THz electric field is sampled using a standard electrooptical detection scheme employing a \SI{1}{\milli\meter} thick ZnTe detection crystal \cite{leitenstorfer1999detectors}.
The magnetic response of the system is quantified by the root mean square (RMS) of half the THz signal difference $S$ for positive and negative magnetic fields.

\section{Results}
\begin{figure}[tb]
	\centering
	\includegraphics[width = 7.63 cm]{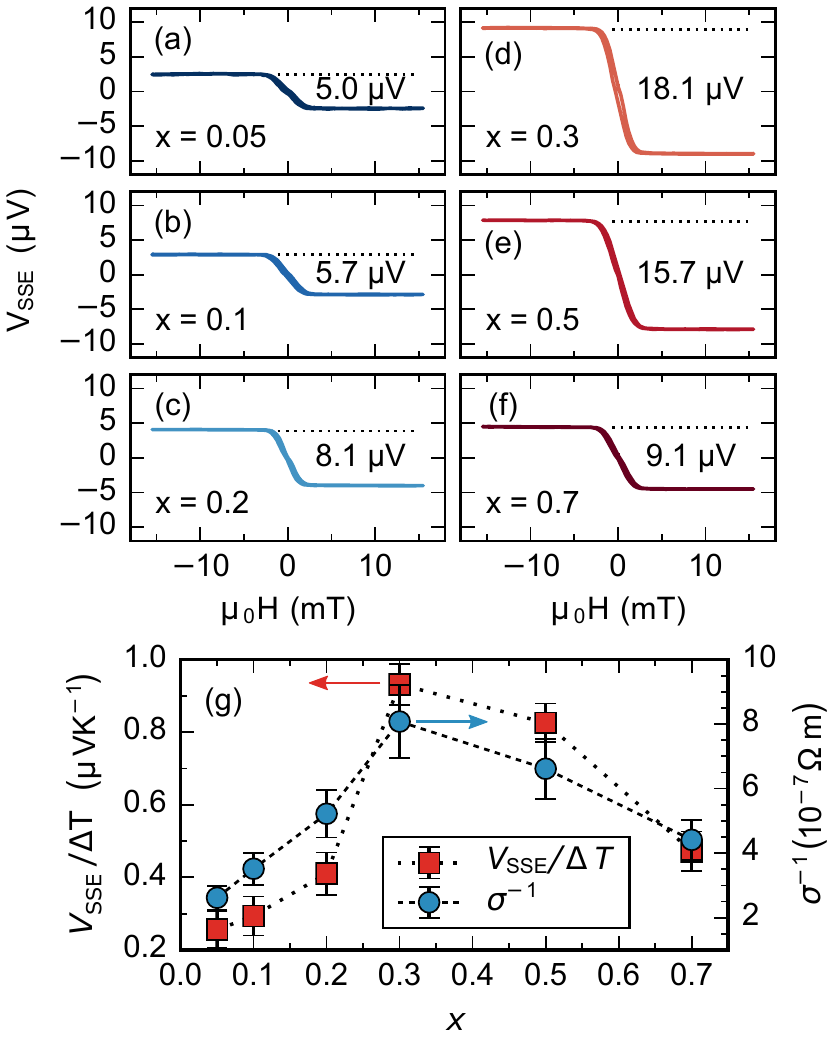}
	\caption{
		(a)-(f)~Measured DC SSE voltage in YIG/Cu$_{1-x}$Ir$_{x}$/AlOx stacks for different Ir
		concentrations~$x$ in ascending order. The temperature difference between sample top and bottom is
		fixed to $\Delta T = \SI{10}{\kelvin}$. (g)~SSE coefficient $V_{\mathrm{SSE}} / \Delta T$ (red
		squares) and resistivity $\sigma^{-1}$ (blue circles) as a function of Ir concentration $x$.}
	\label{fig:sse_overview}
\end{figure}

Figure \ref{fig:sse_overview}(a)-(f) shows DC SSE hysteresis loops measured for
YIG/Cu$_{1-x}$Ir$_{x}$/AlOx multilayers with varying Ir concentration $x$. The temperature
difference between sample top and bottom is fixed to $\Delta T = \SI{10}{\kelvin}$ with a
base temperature of $T = \SI{288.15}{\kelvin}$. In the Cu-rich phase, we observe an increase of the
thermal voltage signal with increasing $x$, exhibiting a maximum at $x = 0.3$. Interestingly, upon further increasing the Ir content $V_{\mathrm{SSE}}$ reduces again.
This behavior is easily visible in Fig.~\ref{fig:sse_overview}(g), in which the SSE coefficient $V_{\mathrm{SSE}} / \Delta T$ is plotted as a function of $x$.
The measured concentration dependence shows that $V_{\mathrm{SSE}}/\Delta T$ exhibits a clear maximum in the range from $x=0.3$ to $0.5$.
Thus, as a first key result the maximum spin Hall effect is obtained for the previously neglected alloying regime beyond the dilute doping.
For comparison, the resistivity $\sigma^{-1}$ of the metal film is also shown in Fig.~\ref{fig:sse_overview}(g).
We see that the resistivity of the Cu$_{1-x}$Ir$_{x}$ layer follows a similar trend as the DC SSE signal.

\begin{figure}[tb]
	\centering
	\includegraphics[width = 7.14 cm]{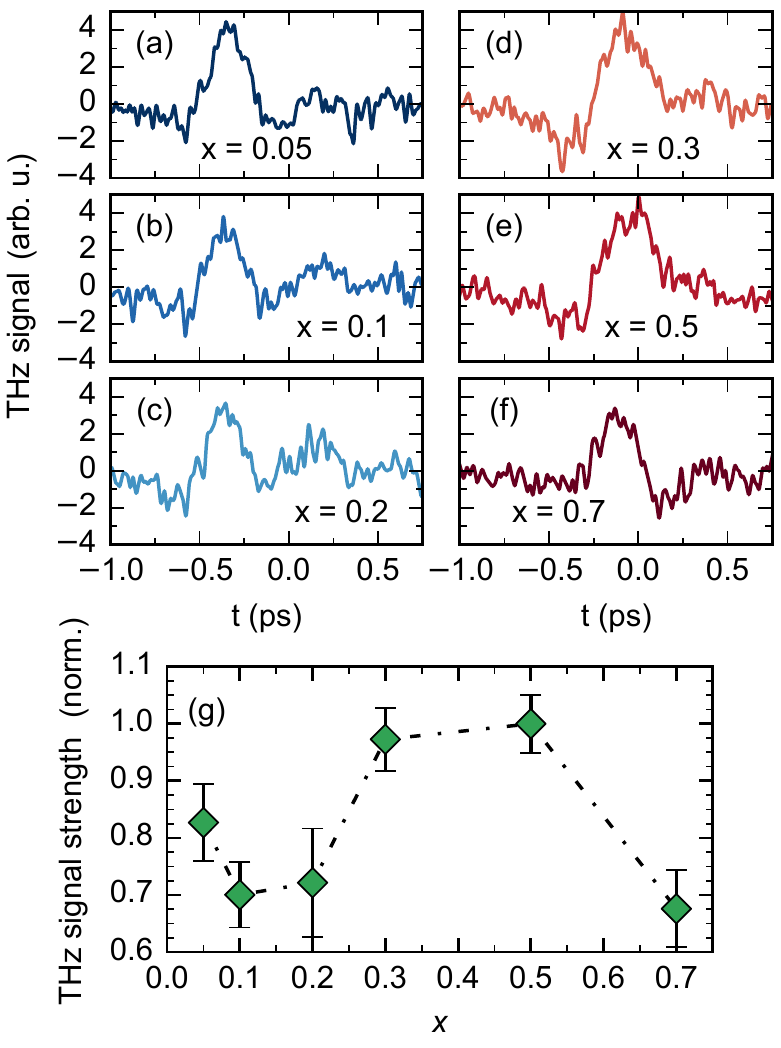}
	\caption{
		(a)-(f)~Signal waveforms (odd in the sample magnetization) of the THz pulses emitted from
		YIG/Cu$_{1-x}$Ir$_{x}$/AlOx stacks for different Ir concentrations~$x$ in ascending order. (g)~THz
		signal strength (RMS) as a function of Ir concentration~$x$.}
	\label{fig:thz_overview}
\end{figure}

Typical THz emission signals from the YIG/Cu$_{1-x}$Ir$_{x}$/AlOx samples are depicted in Figs.~\ref{fig:thz_overview}(a)-(f). The THz transients were low-pass
filtered in the frequency domain with a Gaussian centered at zero frequency and a full width at
half maximum of \SI{20}{\tera\hertz}. The RMS of the THz signal odd in sample magnetization is plotted in
Fig.~\ref{fig:thz_overview}(g) as a function of $x$. After an initial signal drop in the Cu-rich
phase, the THz signal increases with increasing Ir concentration, indicating a signal maximum in
the range between $x = {0.3}$ and ${0.5}$. Further increase of the Ir content leads to a second
reduction of the THz signal strength.

\section{Discussion}
In the following, a direct comparison of the signals obtained from the DC and
the ultrafast THz measurements is established.
To begin with, the emitted THz electric field right behind the sample is described by a generalized
Ohm's law, which in the thin-film limit (film is much thinner than the wavelength and attenuation
length of the THz wave in the sample) is in the frequency domain given by \cite{Seifert2016a}
\begin{equation}
\tilde{E}(\omega) \propto \theta_{\mathrm{SH}}Z(\omega)\int_{0}^{d}\mathrm{d}z \,
j_{\mathrm{s}}(z,\omega) \label{eq:field},
\end{equation}
where $\omega$ is the angular frequency.
The spin-current density $j_{\mathrm{s}}(z,\omega)$ is integrated over the full thickness $d$ of the metal film.
The total impedance $Z(\omega)$ can be understood as the impedance of an equivalent parallel circuit comprising the metal film (Cu$_{1-x}$Ir$_{x}$) and the surrounding substrate (GGG/YIG) and air half-spaces,
\begin{equation}
\frac{1}{Z(\omega)} = \frac{n_1( \omega ) + n_2 ( \omega )}{Z_0} + G(\omega ).
\label{eq:impedance}
\end{equation}
Here, $n_1$ and $n_2 \approx 1$ are the refractive indices of substrate and air, respectively, $Z_0
= \SI{377}{\ohm}$ is the vacuum impedance, and $G(\omega )$ is the THz sheet conductance of the
Cu$_{1-x}$Ir$_{x}$ films. Considering the Drude model and a velocity relaxation rate of
$\SI{28}{\tera\hertz}$ for pure Cu at room temperature as lower boundary \cite{Gall2016}, the values
of $G(\omega )$ vary only slightly over the detected frequency range from 1 to 5~THz (as given by
the ZnTe detector crystal). Therefore, the frequency dependence of the conductance can be
neglected, i.e. $G(\omega) \approx G(\omega=0)$. Importantly, the metal-film conductance ($G\approx
\SI{8e-3}{\per\ohm}$) is much smaller than the shunt conductance ($\left[n_1(\omega) +
n_2(\omega)\right]/Z_0 \approx \SI{4e-2}{\per\ohm}$) for the investigated metal film thickness ($d
= \SI{4}{\nano\meter}$) and can be thus neglected.
Therefore, the Ir-concentration influences the THz emission strength only directly through the ISHE-induced in-plane charge current
flowing inside the NM layer.

The measured DC SSE voltage, on the other hand, is given by an  analogous expression related to the
underlying in-plane charge current by the standard Ohm's law,
\begin{equation}
\frac{V_{\mathrm{SSE}}}{\Delta T} \propto \theta_{\mathrm{SH}}R \int_{0}^{d}\mathrm{d}z \,
j_{\mathrm{s}}(z). \label{eq:VSSE}
\end{equation}
Here, $R$ is the Ohmic resistance of the metal layer between the electrodes, which is inversely proportional
to the metal resistivity $\sigma$, and $j_{\mathrm{s}}(z)$ is the DC spin current density.
Therefore, in contrast to the THz data, the impact of alloying on $V_{\mathrm{SSE}}$ through $\sigma^{-1}$ is significant.
For a direct comparison with the THz measurements, we thus contrast the RMS of the THz signal waveform with the DC SSE current density
$j_{\mathrm{SSE}} = V_{\mathrm{SSE}} \cdot \sigma /\Delta T$.

\begin{figure}[tb]
	\centering
	\includegraphics[width = 7.092 cm]{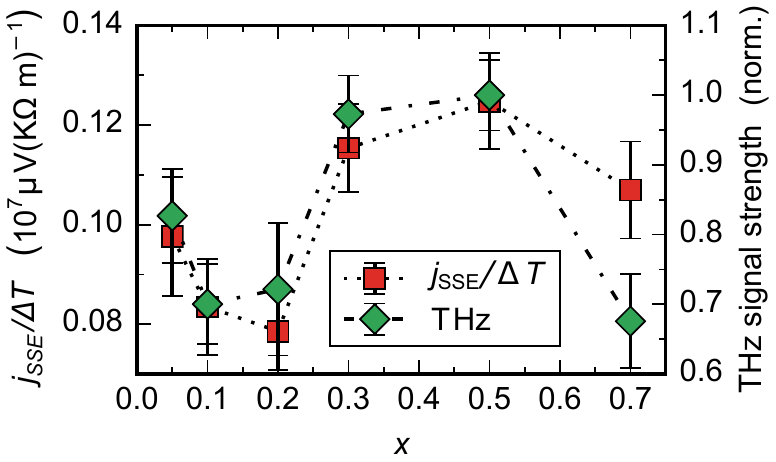}
	\caption{
		Ir concentration dependence of the thermal DC spin current (red squares) and the RMS of the THz
		signal (green diamonds).
	}
	\label{fig:sse_thz_compare}
\end{figure}

In Fig. \ref{fig:sse_thz_compare}, the respective amplitudes are plotted as a function of the Ir
concentration. Remarkably, DC and THz SSE/ISHE measurements exhibit the very same concentration
dependence. This agreement suggests that the ISHE retains its functionality from DC up to THz
frequencies, which vindicates the findings and interpretations of previous experiments \cite{Seifert2016a}. Small discrepancies may originate from a varying optical absorptance of the
near-infrared pump light, which is, however, expected to depend monotonically on $x$ and to only vary by a few percent \cite{Seifert2016a}.
Furthermore, as discussed below, these findings imply that for DC and THz spin currents comparable concentration dependences of spin-relaxation lengths may be expected.


To discuss the concentration dependence of the DC and THz SSE signals (Fig. \ref{fig:sse_thz_compare}), we consider Eqs. (\ref{eq:field}) and (\ref{eq:VSSE}). 
According to these relationships, the THz signal and the SSE voltage normalized by the metal resistivity result from a competition of (i) the spin Hall angle $\theta_{\mathrm{SH}}$ and (ii) the integrated spin-current density $\int_{0}^{d}\mathrm{d}z j_{\mathrm{s}}(z,\omega)$.

At first, we consider the local spin signal minimum at small, increasing Ir concentration $x$ (dilute regime) that appears for both $j_{\mathrm{SSE}}$ and the THz signal.
In fact, with regard to (i) $\theta_{\mathrm{SH}}$ one would expect the opposite behavior as for the dilute regime the skew scattering mechanism has been predicted \cite{Fert2011} and experimentally shown \cite{Niimi2011} to yield the dominant ISHE contribution.
With increasing SOI scattering center density ($\rho_{\mathrm{imp}} \propto \sigma^{-1}$), a linear increase of the spin signal should appear.
In this work, this trend is observed for $V_{\mathrm{SSE}}$ [Fig. \ref{fig:sse_overview}(g)].
The significantly deviating signal shapes of $j_{\mathrm{SSE}}$ and the THz signal, however, suggest that the converted in-plane charge current is notably governed by additional effects.
An explanation can be given by (ii), considering a spatial variation of the spin current density that, as we discuss below, can be influenced by both electron momentum- and spin-relaxation.
The initial electron momenta and spin information of a directional spin current become randomized over length scales characterized by the mean free path $\ell$ and the spin diffusion length $\lambda_{\mathrm{sd}}$, yielding a reduction of the spin current density.
For spin-relaxation, the integrated spin current density is given by \cite{Ando2011}:
\begin{equation}
\int_{0}^{\mathrm{d_{\mathrm{CuIr}}}} dz  j_{\mathrm{s}} (z) \propto \lambda_{\mathrm{sd}} \tanh \left( \frac{d_{\mathrm{CuIr}}}{2\lambda_{\mathrm{sd}}}  \right) j_{\mathrm{s}}^0
\label{eq:jsse}
\end{equation}
with $d_{\mathrm{CuIr}}$ being the thickness of the Cu\textsubscript{1-x}Ir\textsubscript{x} layer.
According to Niimi \textit{et al.} \cite{Niimi2011} the spin-diffusion length $\lambda_{\mathrm{sd}}$ decreases exponentially from $\lambda_{\mathrm{sd}}\approx \SI{30}{\nano\meter}$ for $x=0.01$  to $\lambda_{\mathrm{sd}}\approx \SI{5}{\nano\meter}$ for $x=0.12$.
This exponential decay implies that the integrated spin current density is nearly constant for both small and large $x$, but undergoes a significant decline in the concentration region where $\lambda_{\mathrm{sd}} \approx d_{\mathrm{CuIr}}$.
This effect possibly explains the observed reduction of the signal amplitude from $x = 0.05$ to $x=0.2$.
Furthermore, we interpret the fact that for DC and THz SSE signals similar trends are observed as an indication of similar concentration dependences of $\lambda_{\mathrm{sd}}$ in the distinct DC and THz regimes.
This appears reasonable when considering that spin-dependent scattering rates are of the same order of magnitude as the momentum scattering \cite{Stern2008} (e.g. $\Gamma _{\mathrm{Cu}}^{\mathrm{mom.}} = \SI{1/36}{\femto\second} \approx \SI{28}{\tera\hertz}$ \cite{Gall2016}) and thus above the experimentally covered bandwidth.

In addition to spin-relaxation, the integrated spin current density is influenced by momentum scattering.
As shown in Fig. \ref{fig:sse_overview}, alloying introduces impurities and lattice defects in the dilute phase, such that enhanced momentum scattering rates occur.
Assuming that the latter increase more rapidly than $\theta_{\mathrm{SH}}$, the appearance of the previously unexpected local minimum near $x \approx 0.2$ can be thus explained.

We now focus on the subsequent increase of the spin signal at higher $x$ (concentrated phase).
It can be explained by a further increase of extrinsic ISHE as well as intrinsic ISHE contributions, as pure Ir itself exhibits a sizeable intrinsic spin Hall effect \cite{Tanaka2008,Morota2011}.
A quantitative explanation of the intrinsic ISHE, however, requires knowledge of the electron band structure (obtainable by algorithms based on the tight-binding model \cite{Tanaka2008} or the density functional theory \cite{Guo2008}), which is beyond the scope of this work.
The decrease of $j_{\mathrm{SSE}}$ and the THz Signal at $x=0.7$ may then be ascribed to an increase of atomic order and thus a decrease of the extrinsic ISHE.

In conclusion, we compare the spin-to-charge conversion of steady state and THz spin currents in copper-iridium alloys as a function of the iridium concentration.
We find a clear maximum of the spin Hall effect for alloys of around \SI{40}{\percent} Ir concentration, far beyond the previously probed dilute doping regime.
While the detected DC spin Seebeck voltage exhibits a concentration dependence different from the raw THz signal, very good qualitative agreement between the
DC spin Seebeck current and the THz emission signal is observed, which is well understood within our model for THz emission.
Ultimately, our results show that tuning the spin Hall effect by alloying delivers an unexpected, complex concentration dependence that is equal for spin-to-charge conversion at DC and THz frequencies and allows us to conclude that the large spin Hall effect in CuIr can be used for spintronic applications on ultrafast timescales.

\section{Acknowledgments}
This work was supported by Deutsche Forschungsgemeinschaft (DFG) (SPP 1538 “Spin Caloric Transport”, SFB/TRR 173 "SPIN+X"),
the Graduate School of Excellence Materials Science in Mainz (DFG/GSC 266), and the EU projects IFOX,
NMP3-LA-2012246102, INSPIN FP7-ICT-2013-X 612759, TERAMAG H2020 681917.

\bibliography{bibliography}

\end{document}